\documentclass[aps, prx, reprint, superscriptaddress, showpacs, nofootinbib, longbibliography, floatfix]{revtex4-2}

\usepackage[utf8]{inputenc} 
\usepackage[english]{babel} 
\usepackage{booktabs} 
\usepackage{comment} 
\usepackage{hyperref} 
\usepackage{physics} 
\usepackage[table,xcdraw,dvipsnames]{xcolor} 
\allowdisplaybreaks 

\usepackage{lipsum} 
\usepackage{xfrac} 
\usepackage{multirow} 

\usepackage{amsmath}  
\usepackage{amssymb}  
\usepackage{mathrsfs} 
\usepackage{mathtools} 
\usepackage{amsthm}   
\usepackage{bbm}      
\usepackage{bm}       

\usepackage{indentfirst} 
\usepackage{latexsym}    

\usepackage{verbatim} 
\usepackage{cancel}   
\usepackage{url}      

\usepackage{subcaption} 

\usepackage[font=small, labelfont=bf, format=plain, justification=raggedright, singlelinecheck=false]{caption} 
\usepackage[title, titletoc]{appendix} 

\usepackage{graphicx} 
\graphicspath{{images/}} 
\usepackage{epstopdf} 

\newcommand{\E}[0]{{\mathcal{E}}} 

\newcommand{\caphead}[1]{{\bf #1}} 

\makeatletter
\def\p@subsection{} 
\def\p@subsubsection{} 
\newcommand\footnoteref[1]{\protected@xdef\@thefnmark{\ref{#1}}\@footnotemark} 
\makeatother

\usepackage{soul} 

\usepackage{enumitem}

\newcommand{\sg}[1]{{\color{black}#1}}
\begin{document}

\title{Thermodynamic limitations on fault-tolerant quantum computing}

\author{Mykhailo Bilokur}
\affiliation{Department of Physics Princeton University, Princeton, New Jersey 08544, USA}

\author{Sarang Gopalakrishnan}
\affiliation{Department of Electrical and Computer Engineering, Princeton University, Princeton, New Jersey 08544, USA}

\author{Shayan Majidy}
\email{smajidy@fas.harvard.edu}
\affiliation{Department of Physics, Harvard University, Cambridge, Massachusetts 02138, USA}

\date{\today}

\begin{abstract} 
We investigate the thermodynamic limits on scaling fault-tolerant quantum computers due to heating from quantum error correction (QEC). Quantum computers require error correction, which accounts for 99.9\% of the qubit demand and generates heat through information-erasing processes. This heating increases the error rate, necessitating more rounds of error correction. We introduce a dynamical model that characterizes heat generation and dissipation for arrays of qubits weakly coupled to a refrigerator and identify a dynamical phase transition between two operational regimes: a bounded-error phase, where temperature stabilizes and error rates remain below fault-tolerance thresholds, and an unbounded-error phase, where rising temperatures drive error rates beyond sustainable levels, making fault tolerance infeasible. Applying our model to a superconducting qubit system performing Shor’s algorithm to factor 2048-bit RSA integers, we find that current experimental parameters place the system in the bounded-error phase. Our results indicate that, while inherent heating can become significant, this thermodynamic constraint should not limit scalable fault tolerance if current hardware capabilities are maintained as systems scale.
\end{abstract}

\maketitle

\section{Introduction}\label{sec:int}

\begin{figure}
    \centering
    \includegraphics[width=0.75\linewidth]{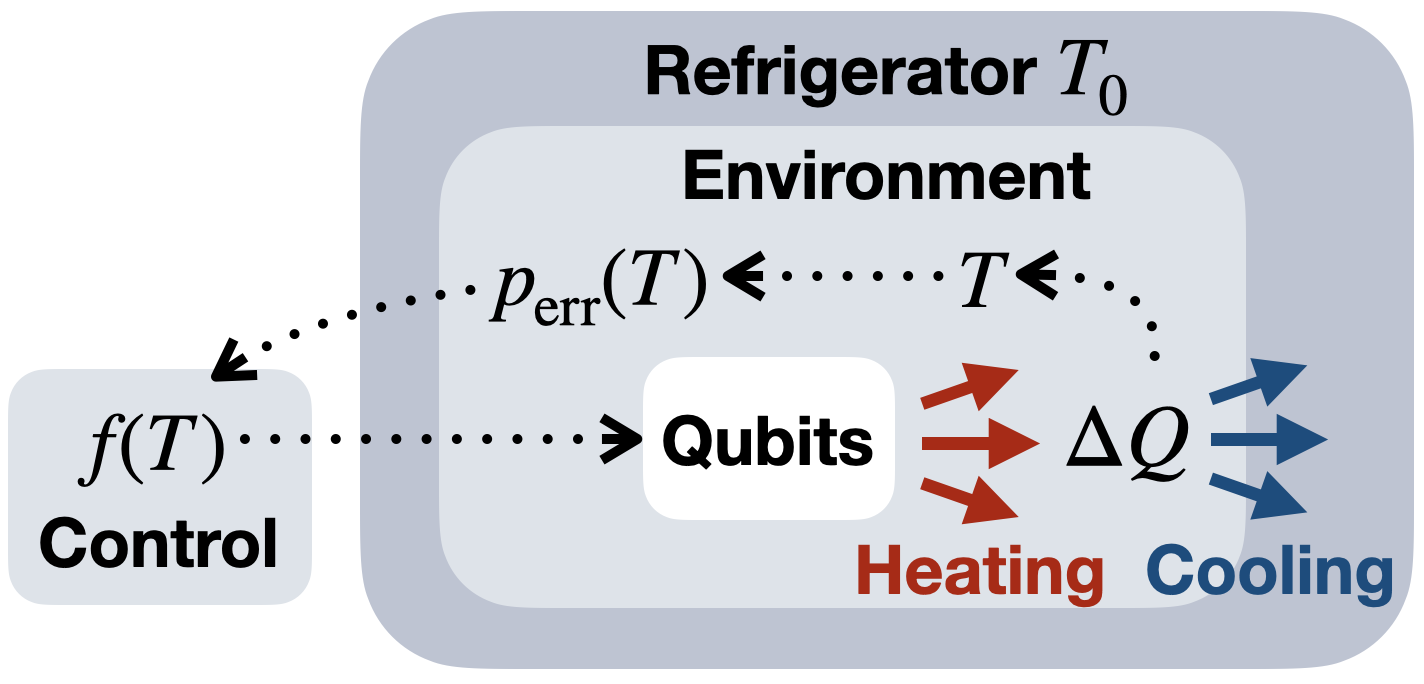}
    \caption{\caphead{Model of heat flow during quantum error correction.} A system of qubits embedded in an environment at temperature $T$, enclosed by a refrigerator at temperature $T_0$. An external device performs periodic QEC rounds to measure and reset the qubits, dissipating heat $\Delta Q$ into the environment. The QEC round frequency $f(p_{\text{err}})$ depends on the error rate $p_{\text{err}}$, which increases with temperature $T$. Each QEC round can raise $T$ and $p_{\text{err}}(T)$, requiring a higher $f(T)$.}
    \label{fig:model}
\end{figure}

Quantum error correction (QEC) is inherently a dissipative process---in which the system is made to forget which error happened---and therefore generates heat in accordance with Landauer's principle~\cite{landauer1961irreversibility}. Standard QEC protocols involve measuring error syndromes using ancilla qubits, performing a recovery operation conditional on the syndromes, and then erasing the ancillas in preparation for the next round of QEC. This last step generates heat. In present-day quantum computing devices, this heating occurs in classical control devices that are physically separated from the qubits~\cite{majidy2024building}. However, a truly scalable architecture involving many thousands of logical qubits would require QEC to be done ``on-chip,'' in proximity to the qubits. In these conditions, the heat that each erasure generates will warm up the environment for nearby system qubits, and increase their error rates. In the absence of cooling mechanisms, Landauer heating then leads to a runaway process: QEC generates heat, increasing the error rate, necessitating an increase in the frequency of QEC, which accelerates heating, and so on until the error rate crosses the error correction threshold (Fig.~\ref{fig:model}).

To overcome this heating, a quantum computer needs to be continuously cooled. The main result of this paper is that fault tolerance is only attainable when the cooling rate exceeds a threshold. Our primary result identifies a dynamical phase transition between two regimes: in the \textit{bounded-error phase}, the temperature around ancillary qubits stabilizes, maintaining a steady error rate below the error threshold and enabling scalable quantum computing. The temperature rises uncontrollably in the \textit{unbounded-error phase}, driving $p_{\rm err}$ to pass the threshold and making fault tolerance unattainable. This thermodynamic threshold applies to any form of autonomous open-system dynamics that protects a quantum memory in its steady state~\cite{harrington2004analysis, herold2015cellular, PhysRevLett.123.020501, chirame2024stabilizing}. In previous work, such dynamical processes were presented as quantum channels or Lindblad master equations; our result implies that physical implementations of such channels must extract energy from the system fast enough to counteract Landauer heating.

This phase transition raises a critical question: to what extent can quantum computers scale before reaching an unbounded-error regime? To investigate this, we apply our model to a specific physical system and algorithm—a superconducting qubit architecture designed to factor 2048-bit RSA integers using Shor’s algorithm~\cite{shor1999polynomial}, with parameters based on modern superconducting devices~\cite{kjaergaard2020superconducting}. Our goal is to provide an order-of-magnitude estimate based on transparent and well-supported assumptions, not to give a precise prediction of the future. We find that if current experimental parameters are maintained as systems scale to $10^{7}$ qubits, the system will be in the bounded-error phase for this task.

This manuscript is organized as follows: Sec.~\ref{sec:bac} reviews Landauer's principle and its implications for fault-tolerant quantum computing. Sec.~\ref{sec:mod} describes our numerical model, Sec.~\ref{sec:pha} presents the dynamical phase transition, Sec.~\ref{sec:phy} applies this model with experimental parameters, and Sec.~\ref{sec:dis} summarizes results and outlines future research directions.

\section{Background}\label{sec:bac}

For completeness, we briefly review Landauer's principle and essential elements of the threshold theorem for fault-tolerant quantum computing, starting with key notation. The von Neumann entropy of a quantum state $\rho$ is defined as $ S(\rho):= -\tr[\rho \log \rho] $. For a bipartite system $ \rho_{AB} $, mutual information is $ I(\rho_A : \rho_B) := S(\rho_A) + S(\rho_B) - S(\rho_{AB}) $, with $ \rho_x = \tr_{\overline{x}}[\rho_{AB}] $ for $ x = A, B $. The quantum relative entropy between states $ \sigma $ and $ \rho $ is $ D(\sigma || \rho) := \tr[\sigma \log \sigma] - \tr[\sigma \log \rho] $.

Landauer’s principle is derived by considering a closed system with a subsystem $ S $ initially in state $ \rho_S $ and an environment $ \mathcal{E} $ in thermal state $ \rho_{\mathcal{E}} = \frac{e^{-\beta H}}{\tr(e^{-\beta H})} $, where $ H $ is the Hamiltonian and $ \beta $ is the inverse temperature. Initially, the total state is $ \rho_{S\E} = \rho_S \otimes \rho_{\E} $, which evolves unitarily to $ \rho'_{S\E} = U \rho_{S\E} U^\dagger $. The reduced states are $ \rho'_x := \tr_{\overline{x}}[\rho'_{S\E}] $ for $ x = \{S, \mathcal{E}\} $.

Using $ S(\rho) $ and $ I(\rho_A : \rho_B) $, we find that total entropy production is non-negative,
\begin{equation}
S(\rho'_S) - S(\rho_S) + S(\rho'_{\mathcal{E}}) - S(\rho_{\mathcal{E}}) = I(\rho'_S : \rho'_{\mathcal{E}}) \geq 0,
\end{equation}
which follows from the mutual information being non-negative. Defining $ \Delta S_x := S(\rho_x) - S(\rho'_x) $ as the entropy decrease for $ x = \{S, \mathcal{E}\} $ and heat transfer to the environment as $ \Delta Q = \tr[H(\rho'_{\mathcal{E}} - \rho_{\mathcal{E}})] $, we establish~\cite{reeb2014improved}:
\begin{equation}
\beta \Delta Q = \Delta S_S + I(\rho'_S : \rho'_{\mathcal{E}}) + D(\rho'_{\mathcal{E}} || \rho_{\mathcal{E}}) \geq \Delta S_S.
\end{equation}
This inequality encapsulates Landauer’s principle: reducing a system’s entropy requires a corresponding heat transfer to the environment~\cite{landauer1961irreversibility}.

Landauer's principle predicts that QEC will cause heat dissipation. In each QEC round, ancillary qubits entangle with computational qubits, are measured to extract errors, and are then reset. This reset lowers the ancillas' entropy, releasing heat into the environment.

The threshold theorem in fault-tolerant quantum computing states that computations can be sustained if $ p_{\rm err} $ remains below a threshold $ p_{\rm th} $ that depends on the error-correcting code~\cite{aharonov1997fault, knill1998resilient}. If $ p_{\rm err}(T) $ exceeds this threshold, errors accumulate faster than they can be corrected, breaking fault tolerance.

Since each QEC round generates heat during ancilla measurements and resets, effective cooling is essential to keep $ p_{\rm err}(T) $ below $ p_{\rm th} $. Insufficient cooling could cause cumulative heat buildup, raising $ T $ and pushing error rates beyond sustainable levels. Thus, understanding QEC's thermal effects is vital for designing hardware that supports sustained fault-tolerant quantum computing.

\section{Model}\label{sec:mod}

\begin{figure}
    \centering
    \includegraphics[width=0.55\linewidth]{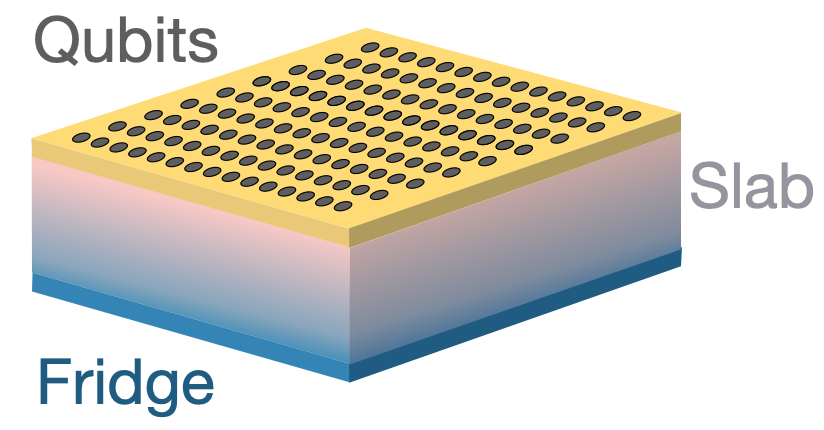}
    \caption{\caphead{Representation of the physical setup.} The model consists of a slab geometry with a two-dimensional array of qubits at the top, where erasure processes generate heat. A fridge is coupled to the bottom for cooling. Heat transport occurs via diffusion through the slab, with the local temperature at the top influencing the error correction rate. This simplified one-dimensional heat flow model assumes uniform qubit parameters and serves as a representative framework for arrays of superconducting qubits. 
    }
    \label{fig:lattice}
\end{figure}

\sg{We first outline the physical setup we will consider. It consists of a slab geometry with a two-dimensional array of qubits at the top of the slab and a fridge coupled to the bottom of the slab. We assume that heat is generated by erasure processes occurring near the top of the slab (where the qubits are), and is transported by diffusion. The error correction rate is a function of the local temperature at the top of the slab. This is a plausible toy model for arrays of superconducting qubits; we will comment later on its extension to other types of platforms. We will treat the simplest case, in which all qubits have the same parameters, so the heat flow is purely one-dimensional, i.e., from the qubits to the fridge.}

\sg{We now present the specifics of the model.} The model is governed by three temperature-dependent parameters: the rate\sg{s at which} heat is generated, \sg{transported}, and removed. Dimensionful constants are included to ensure transparency and avoid hidden temperature dependencies, clarifying the rate of heat production and removal. A time-dependent simulation is necessary to capture the feedback loop in QEC, where increasing QEC demand can generate more heat.

\sg{Ignoring spatial variation in the transverse directions we can model the environment as a finite one-dimensional segment of diffusive material, with (physical) qubits at one end and the fridge at the other, separated by a distance $L$. (Technically the qubits that are not measured and reset are not relevant to our model, but this distinction will not matter.) Positions along the transverse direction are labelled $x$ and each position corresponds to a temperature $T(x)$. Our model generalizes to the case where each qubit is surrounded by a higher-dimensional environment, but the one-dimensional slab geometry is the most natural, and we will focus on that in what follows.} 

The system evolves in discrete time steps $ \Delta t $. We model temperature dynamics by updating the temperature $ \Delta T_{\vec{r}}(t + \Delta t) $ at each site. The temperature update includes three terms: heating from QEC, diffusion through the lattice, and cooling by the refrigerator. In each time step, heat diffuses across the lattice and the boundary sites are cooled. QEC deposits heat intermittently. The time step $ \Delta t $ is initially much shorter than the interval between QEC rounds, allowing heat diffusion to appear continuous relative to the QEC cycle.

To model temperature dynamics, we quantify the net heat added or removed at each lattice site, normalized by the site’s heat capacity. We assume that our quantum computer operates below the Debye temperature, $\Theta_{\rm D}$, allowing us to apply the low-temperature Debye approximation for heat capacity. This assumption is valid across various types of quantum hardware, such as superconducting qubits. Furthermore, we enter the unbounded error phase at temperatures significantly below $\Theta_{\rm D}$, ensuring this approximation remains applicable. For instance, the Debye temperature for silicon is approximately $636 \text{K}$. When $T \ll \Theta_{\rm D}$, the Debye model gives the heat capacity as:
\begin{equation}\label{eq:heatcapacity}
    C_{\rm HC} (T_{\vec r}) \approx \frac{12 \pi^4}{5} \frac{Nk_B}{\Theta_{\rm D}^3} T_{\vec r}^3 =: AT_{\vec r}^3,
\end{equation}
where $N$ is the number of atoms, $k_B$ is the Boltzmann constant, $C_{\rm HC}$ denotes the heat capacity, and $A$ is a defined proportionality constant. This expression provides the bulk heat capacity in joules per kelvin.

Our first term captures the heating. We assume the Landauer-generated heat is evenly distributed among the $ 2d $ neighbouring lattice sites around the $n_a$ qubits. The inverse QEC frequency, $ f(T_{\vec{r}}) $, sets the time steps $ \Delta t $ between successive heat depositions. We define a binary function $ \mathcal{Q}[f(T_{\vec{r}})] $, equal to 1 when QEC occurs in a time step and 0 otherwise. The temperature change from QEC heating at site $ \vec{r} $ is:
\begin{equation}\label{eq:Landaurheating}
   \Delta T^{(1)}_{\vec{r}}(t + \Delta t) = \frac{1}{C_{\rm HC}}
    \frac{n_a \ln(2)}{2d} k_B T_{\vec{r}} \delta_{\vec{r},\vec{r}_1} \mathcal{Q}[f(T_{\vec{r}})],
\end{equation}
where $ \delta_{\vec{r},\vec{r}_1} $ is the Kronecker delta, equal to 1 for nearest neighbours $ \vec{r} = \vec{r}_1 $ and 0 otherwise.

The second term represents heat diffusion, modelled using the discrete form of Fourier’s law (see Sec.~3 of Ref.~\cite{challamel2016nonlocal}). The temperature change at site $ \vec{r} $ due to diffusion is:
\begin{equation}\label{eq:Diffusion1}
      \Delta T^{(2)}_{\vec{r}}(t + \Delta t) = \frac{1}{C_{\rm HC}} \frac{\kappa V \Delta t}{a^2}
     \sum_{\abs{\vec r - \vec r'} = 1}(T_{\vec r'} - T_{\vec r}),
\end{equation}
where $ a $ is the lattice spacing, $ \kappa $ is thermal conductivity, and $ V $ is the samples volume. Assuming the phonons in the substrate behave diffusely (see Eq.~(3.25) of Ref.~\cite{pobell2007matter}), thermal conductivity $ \kappa $ is:
\begin{equation}
    \kappa = \frac{1}{3} C_{\rm HC} \frac{\Lambda \bar{c}}{V},
\end{equation}
where $ \Lambda $ is phonon mean free path and $ \bar{c} $ is average sound speed. While $ \Lambda $ may depend on temperature, it is ultimately bounded above by the system's smallest dimension. Consequently, as discussed in Sec.~\ref{sub:parameters}, we consider it to be constant in the cases studied here.

The third term represents heat removal via peripheral lattice sites in contact with a refrigerator. The temperature dependence of the cooling rate will depend on the cooling method used. Here, we assume a dilution refrigerator, but this model can be adapted for other cooling methods.

The total cooling capacity of a dilution refrigerator is given by $ \dot{Q} = 84 \dot{n}_3 T_F^2 $~\cite{lounasmaa1974experimental}, where $ T_F $ is the fridge temperature and $ \dot{n}_3 $ is the helium-3 flow rate.\footnote{Although the right-hand side suggests otherwise, $ \dot{Q} $ has units of J/s, as clarified in Eq.~3.23 of Ref.\cite{lounasmaa1974experimental}.} Directly adding this cooling term would lead to an unrealistic model, allowing the system to cool indefinitely, even to temperatures below zero if no QEC heating occurs. Additionally, in practice, the fridge temperature is not entirely fixed but increases with the system's temperature.

The first issue is resolved by considering the steady-state heat load, which represents the constant heat that must be removed to maintain a stable temperature. This steady-state heat load is $84 \dot{n}3 T_0^2$, where $T_0$ is the base temperature. To address the second issue, we note that the boundary lattice site is in thermal equilibrium with the fridge so that $T_F = T{\vec{r}_L}$. Furthermore, we divide this total cooling capacity by the number of sites being cooled, $n_c$. Incorporating these considerations, the temperature change due to cooling at site $ \vec{r} $ becomes: 
\begin{align}\label{eq:CoolTerm}
    \Delta T^{(3)}_{\vec{r}}(t + \Delta t) = \frac{1}{C_{\rm HC}(T_{\vec{r}})} \frac{84 \dot{n}_3}{n_c} \Delta t n_{\vec{r}} (T_0^2 - T_{\vec{r}_L}^2),
\end{align}
where $ n_{\vec{r}} $ is the number of refrigerator sites adjacent to $ \vec{r}_L $.

Summing the contributions from heating, diffusion, and cooling, we derive the temperature at site $\vec{r}$ after one time step:
\begin{widetext}
\begin{align}\label{eq:TempMap}
    T_{\vec r}(t+\Delta t) = T_{\vec r}(t) + \frac{\alpha }{T_{\vec r}^2}  \delta_{\vec{r},\vec{r}_1}  \mathcal{Q}[f(T_{\vec{r}})] +  \delta \Big( \sum_{\abs{\vec r - \vec r'} = 1}T_{\vec r'}-T_{\vec r} \Big) +  \frac{\gamma}{T_{\vec r}^3}n_{\vec{r}}(T_0^2-T_{\vec{r}_L}^2).
\end{align}
\end{widetext}
Here we define the coefficients $\alpha \coloneqq \frac{n_a k_B \ln(2)}{2dA}$, $\gamma \coloneqq \tfrac{84 \dot{n}_3 \Delta t}{An_c} $, and $\delta \coloneqq \tfrac{\Lambda \bar{c}\Delta t}{3a^2}$, representing the contributions from QEC heating, refrigerator cooling, and heat diffusion, respectively.

To complete the model, we need an expression for the QEC frequency $ f(T_{\vec{r}_1}) $ in terms of the error probability $ p_{\rm err} $. This function must satisfy:
\begin{itemize}[noitemsep]
    \item $ f(T_{\vec{r}_1}) \rightarrow 0 $ as $ p_{\rm err} \rightarrow 0 $ (no errors, no QEC required).
    \item $ f(T_{\vec{r}_1}) \rightarrow \infty $ as $ p_{\rm err} \rightarrow p_{\rm th} $ (errors exceed threshold, QEC insufficient).
    \item $ f(T_{\vec{r}_1}) $ increases monotonically with $ p_{\rm err} $.
\end{itemize}
To our knowledge, the precise form of this function remains uncertain. We assume that the simplest form satisfying these conditions will suffice. Let $p_f$ denote the logical failure probability, determined by the code parameters. We define $f(T_{\vec{r}1}) = \left(\frac{p_f}{1 - p_f}\right)^{c_f}$, ensuring all three conditions are met. Here, $c_f$ is a tunable parameter that adjusts how progressively the QEC rate increases. A large $c_f$ implies minimal QEC intervention until the error likelihood is substantial, at which point QEC is applied rapidly. In contrast, a small $c_f$ implies a more gradual increase in the QEC rate as the error probability rises. The specific relationship between $p_f$ and $p{\rm err}$ depends on the chosen QEC code.

\section{Phase Transition in Error Rates and Fault-Tolerance}\label{sec:pha}

In this section, we analyze the model presented in Sec.~\ref{sec:mod} by varying the parameters $\alpha$, $\delta$, and $\gamma$ to explore its operational phases. Through this analysis, we identify a dynamical phase transition that separates two distinct operational regimes of a quantum computer: the bounded-error and unbounded-error phases, which we describe in detail below.

In the bounded-error phase, the temperature surrounding the ancilla qubits stabilizes, keeping the error probability $ p_{\rm err} $ below the fault-tolerance threshold $ p_{\rm th} $. This stability allows for fault-tolerant quantum computing. If $ p_{\rm err} $ stabilizes, it must do so below $ p_{\rm th} $. Starting with an initial $ p_{\rm err} < p_{\rm th} $, any stabilization above $ p_{\rm th} $ would mean the system crossed the threshold, causing the QEC frequency $ f(T_{\vec{r}_1}) $ to diverge, leading to a temperature spike. Thus, any plateau in $ p_{\rm err} $ must be within the fault-tolerant regime.

In contrast, in the unbounded-error phase, the temperature near the qubits continues to rise indefinitely, pushing $ p_{\rm err} $ beyond $ p_{\rm th} $. Under these high error rates, error correction becomes ineffective, making fault-tolerant quantum computing impossible. Before reaching this phase, the assumption that heat deposition is slower than diffusion breaks down, at a failure time $ \tau $. By this point, the temperature increase is so rapid that $\tau$ is effectively the time it takes for the computation to fail. 

\begin{figure}
    \centering
    \includegraphics[width=\columnwidth]{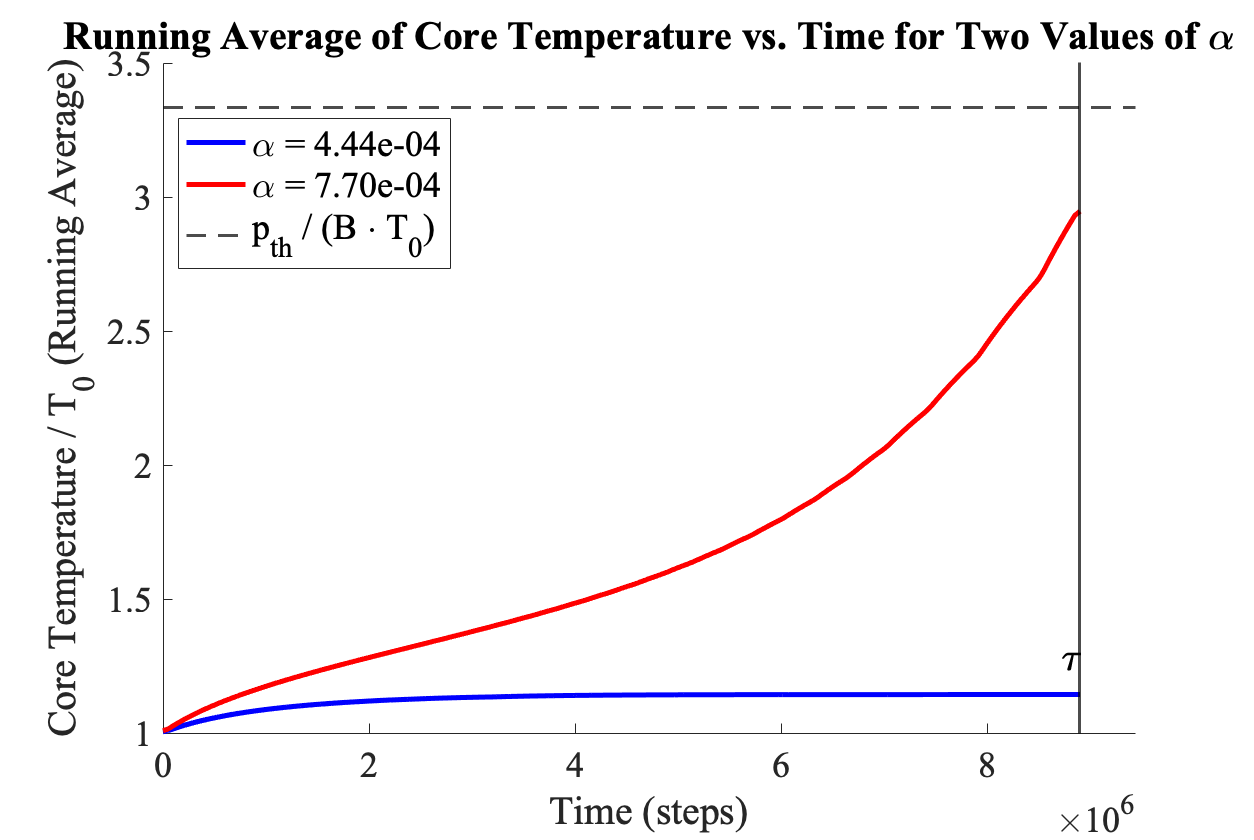}
    \caption{\caphead{Dynamical behaviour of the two phases}
     The temperature immediately surrounding the qubits, $T_{\vec{r}_1}(t)$, is plotted against time for different values of the cooling coefficient $\gamma$, with fixed $\alpha$ and $\delta$. The dashed line represents the temperature corresponding to the error threshold $p_{\rm th}$. In the bounded-error phase (blue curves), the temperature stabilizes over time. In the unbounded error phase (red curves), the temperature continues to rise, eventually exceeding the threshold. The time $\tau$ marks the point where the rate of heat addition becomes comparable to the rate of heat diffusion. The numerics stop at this point. Fitting the curve and extrapolating, we find the curve will cross the threshold barrier shortly after $\tau$.
    }
    \label{fig:example}
\end{figure}

Figure~\ref{fig:example} illustrates the difference in these phases' dynamical behaviour. The plots show temperature $ T_{\vec{r}_1}(t) $ at neighbouring sites over time for different cooling coefficients $ \gamma $, with $ \alpha $ and $ \delta $ fixed. In both phases, temperature initially rises rapidly due to low initial heat capacity, as described by the Debye heat capacity approximation in Eq.~\eqref{eq:heatcapacity}. The heat capacity grows cubically with temperature, slowing the rise as the system absorbs more heat. In the bounded-error phase, temperature stabilizes below the critical threshold, keeping $ p_{\rm err} $ manageable. In the unbounded-error phase, the temperature continues to rise, causing $ p_{\rm err} $ to increase uncontrollably.

\begin{figure}[htbp]
    \centering
    \includegraphics[width=\columnwidth]{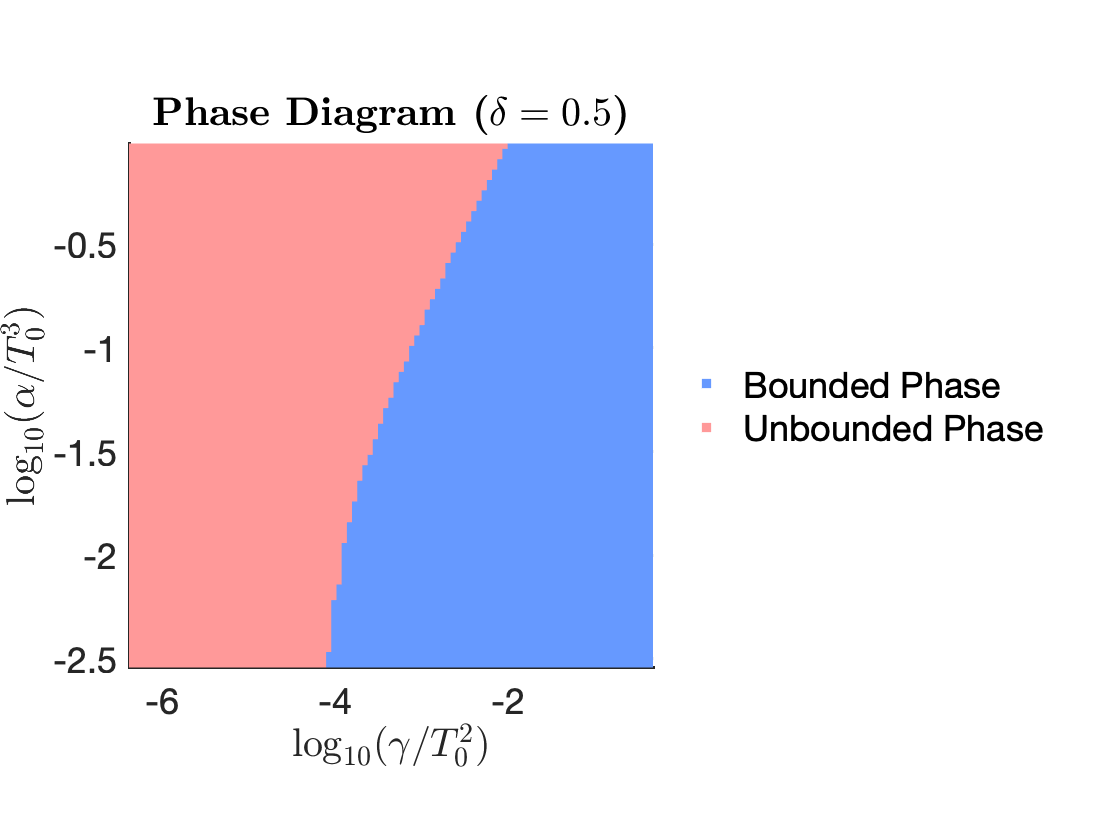}
    \caption{\caphead{Phase diagram}  
     The phase diagram plots the heating coefficient $\alpha$ against the cooling coefficient $\gamma$ for fixed values of the diffusion coefficient $\delta$. The blue region indicates the bounded-error phase, and the red region indicates the unbounded-error phase. The terms are called by $T_0^3$ and $T_0^2$, respectively, to ensure the term inside the log is unitless. Note that the Shor point is not included in this diagram but is shown in Figure \ref{fig:numerical_Estimates}.
    }
    \label{fig:phase_diagram}
\end{figure}

\begin{figure}
    \centering
    \includegraphics[width=\columnwidth]{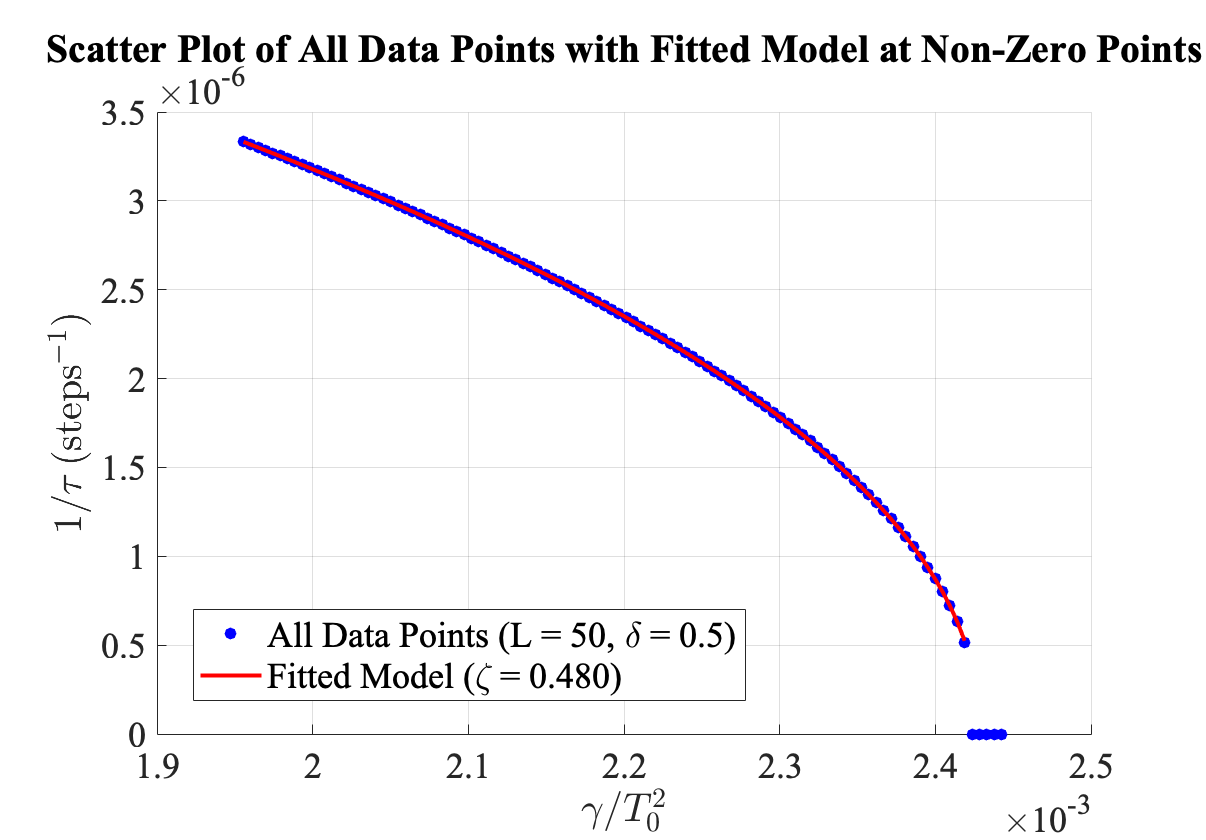}
    \caption{\caphead{Sharpness of the phase transition.} 
    The cooling coefficient $\gamma$ is plotted against $1/\tau$, where $\tau$ is effectively the time at which the error probability $p_{\rm err}$ exceeds the threshold $p_{\rm th}$. A finite value of $1/\tau$ indicates the unbounded-error phase, while $1/\tau = 0$ corresponds to the bounded-error phase with infinite $\tau$. The plot shows a sharp transition between the two phases. We fix $\alpha = 10^{-6}K^3$ and $\delta = 0.5$.
    }
    \label{fig:phasetransition}
\end{figure}

Figure~\ref{fig:phase_diagram} contains phase diagrams constructed by plotting the heating coefficient $ \alpha $ against the cooling coefficient $ \gamma $ for fixed values of the diffusion coefficient $ \delta $. To highlight the sharpness of the transition between these phases, we also analyze how the cooling capacity $ \gamma $ affects the failure time $ \tau $. Figure~\ref{fig:phasetransition} shows $ \gamma $ versus the inverse failure time, $ 1/\tau $, distinguishing between a phase with finite $ \tau $ (unbounded-error phase) and one with infinite $ \tau $ (bounded-error phase). Numerical fits in Figure~\ref{fig:phasetransition} indicate a critical exponent $ \zeta \approx 1/2 $ characterizing the phase transition.

\section{Thermodynamic Limits for Realistic Hardware}\label{sec:phy}

Regardless of reductions in other sources of heat, the heat generated by Landauer’s principle is inevitable. As quantum computers scale, this heat will eventually present a challenge. In this subsection, we estimate the point at which Landauer heating becomes problematic. To achieve this, we first specify a physical hardware platform along with its relevant parameters and select a target computational task, as detailed in Sec.~\ref{sub:parameters}. Next, we numerically analyze this setup in Sec.~\ref{sub:boundestimates}, assessing whether a typical superconducting qubit system could feasibly run Shor’s algorithm to factor a 2048-bit RSA integer within acceptable error bounds, and estimating how long large-scale quantum computations might proceed before qubit reset-induced cooling limitations become problematic.

\subsection{Parameter Selection}\label{sub:parameters}

The task at hand is factoring 2048-bit RSA integers using Shor's algorithm, providing a practical benchmark for defining “large-scale” quantum computing. Various estimates of the required circuit size and depth for this task have been proposed~\cite{gidney2021factor, vedral1996quantum, beauregard2002circuit, fowler2012surface, haner2016factoring, roetteler2017quantum, zalka1998fast, gheorghiu2019benchmarking, jones2012layered, campbell2017quantum, van2010distributed}. Even with fixed error rates, estimating the number of qubits and circuit depths for this task is challenging. Early estimates suggested approximately 6.5 billion physical qubits running over 410 days, while recent advances have reduced this to around 20 million qubits operating for 8 hours. These values depend on factors such as the physical error rate, qubit connectivity, choice of encoding, code cycle times, and decoding rates. Since our error rate is dynamic, we adopt the following approach: we assume a requirement of at least $10^7$ physical qubits from Ref.~\cite{gidney2021factor} and analyze the system’s steady-state behaviour. Following Ref.~\cite{gidney2021factor}, we assume surface code error correction with a code distance of $27$.

To set our physical parameters, we focus on a superconducting qubit architecture, specifically transmon qubits on a silicon substrate. A potentially important factor in this study is the size of the silicon substrate. Current devices use substrates with dimensions of approximately 10 mm $\times$ 10 mm $\times$ 0.5 mm, accommodating 50–100 qubits. The number of qubits on such chips has grown without an appreciable growth in the chips, so it is plausible more qubits can fit on current chips. However, when the number of qubits exceeds the capacity of a single chip, arbitrarily increasing chip size is not feasible due to yield loss during fabrication. Therefore, solutions are more likely to involve interconnected chips rather than larger individual chips.

The increased scale could potentially impact our model in two ways. First, by influencing the heat capacity, which depends on the total number of atoms in the superstate. Although heat capacity affects both heating and cooling equally—making it unlikely to become a major factor—we still estimate this value. As the number of qubits scales by a factor of $10^5$, the amount of silicon associated with the qubits is expected to increase proportionally, requiring approximately $4.15 \times 10^3$ moles of silicon, or roughly $N \approx 2.5 \times 10^{27}$ atoms. Second, scaling could in principle affect the phonon mean free path, limited by the smallest dimension of the silicon wafer. The thickness of the silicon wafer does not necessarily need to increase with the qubit count and can remain at 0.5 mm.

We begin by evaluating the diffusion coefficient, $\delta = \frac{\Lambda \bar{c} \Delta t}{3 a^2}$. For silicon, the average speed of sound across all polarizations is $\bar{c} = 5718$ m/s \cite{hauer2020chip}. In weakly doped silicon at temperatures around 1 K or lower, the mean free path is approximately 1 mm (see Fig. 5 of Ref.~\cite{weber1991transport}). Since our chips are 0.5 mm thick, the phonon mean free path is primarily limited by the chip thickness. We assume $\Lambda = 0.5$ mm and treat it as temperature-independent within our operational regime.

To conduct a numerical model with 50 lattice sites, we set the lattice spacing $a$ such that $a \times L \approx 0.05$ mm for $L = 50$, yielding $a = 1\, \mu$m. This choice enables accurate modelling of local temperature gradients while keeping computational demands manageable for large simulations \cite{clarke2008superconducting}. To ensure diffusion appears continuous relative to quantum error correction (QEC), we select a small time step, $\Delta t$, to capture the rapid initial temperature changes associated with low heat capacity at ultra-low temperatures. This also prevents numerical errors that could arise from larger time steps in explicit methods. To satisfy the Courant–Friedrichs–Lewy stability condition for the heat equation \cite{weber1991transport}, we require $\Delta t \leq \frac{a^2}{2\mathcal{J}} \approx 0.5$ ps, where $ \mathcal{J} = \kappa V / C_{\rm HC} = \Lambda \bar{c}/3 \approx 0.95$m$^2$s$^{-1}$ is the thermal diffusivity. We set $\Delta t = 0.526$ ps and compute:
\begin{align}
    \delta = \frac{\Lambda \bar{c}\Delta t}{3 a^2} \approx 0.5.
\end{align}
Here $\delta$ is dimensionless.

The proportionality constant $ A $ related to the Debye heat capacity is defined as $ A = \frac{12 \pi^4}{5} \frac{N k_B}{\Theta_{\rm D}^3} $. Since $ A $ scales with both heating and cooling rates, small variations are unlikely to significantly impact the overall dynamics. For silicon, the Debye temperature is approximately $\Theta_{\rm D} \approx 636$ K \cite{keesom1959specific}. Using this value along with our earlier estimate for $ N $, we find:
\begin{align}
    A = \frac{12 \pi^4}{5} \frac{Nk_B}{\Theta_{\rm D}^3} \approx 3.14 \times 10^{-2} {\rm JK}^{-4}.
\end{align}

The cooling parameter, $\gamma$, is defined as $\gamma = \frac{84 \dot{n}_3 \Delta t}{A}$. For realistic values, we use parameters from a BlueFors LD dilution refrigerator, a widely used industry standard \cite{Bluefors2024}, where $84 \dot{n}_3 \approx 0.04 \, {\rm W K}^{-2}$. We take the base temperature to be $T_0 = 10$mK. Due to the symmetry of the $d = 1$ setup, we place the qubits at one end of the lattice rather than in the center, which alters our heating and cooling terms by a factor of two, represented here by $n_c = 1$. Substituting these values, we find:
\begin{align}
    \gamma = \frac{84 \dot{n}_3 \Delta t}{An_c} \approx 6.7 \times 10^{-13} {\rm K}^2.
\end{align}
The heating parameter, $\alpha$, is given by $\alpha = \frac{n_a k_B \ln(2)}{2d A}$. For our $d = 1$ system, we have $n_a = 20 \times 10^6$ qubits. Given our setup with qubits at one end of the lattice, we remove the factor of 2 in the denominator. Substituting these values, we get:
\begin{equation}
    \alpha =  \frac{n_a k_B \ln(2)}{dA} \approx 8.79 \times 10^{-15} {\rm K}^3.
\end{equation}

Next, we fix the function $ f(T_{\vec{r}_1}) $, which relates the error correction frequency to the error probability $ p_{\rm err} $. Since our qubit estimates are based on the surface code, we use a threshold error rate of $ p_{\rm th} = 1\% $ \cite{dennis2002topological}. The probability of logical failure in the surface code scales as $ p_f = \left(p_{\rm err}/p_{\rm th}\right)^{d_c/2} $~\cite{gottesman2013fault}, where $ d_c $ is the code distance. Following Ref.~\cite{gidney2021factor}, we use $ d_c = 27 $. Recall to relate the failure probability to the error correction frequency, we set $ f(T_{\vec{r}1}) = \left(\frac{p_f}{1 - p_f}\right)^{c_f} $. Here we choose $ c_f = 1/4 $, so that the frequency of QEC increases gradually with the error rate. This approach prevents scenarios where QEC is essentially absent or suddenly applied at a very high rate.

Various studies have related environmental temperature to error probabilities in superconducting qubit systems \cite{martinis2005decoherence, sun2012measurements, paladino20141, lenander2011measurement, lisenfeld2007temperature}. However, to the best of our knowledge, a model of $p_{\rm err}$ as a function of temperature does not exist. In general, we infer that $ p_{\rm err} $ can be approximated by three primary contributions. The first is a base error probability, $ p_0 $, which is temperature-independent and arises from inherent decoherence mechanisms. The second term accounts for quasiparticle errors, described by $ A e^{-\Delta / k_B T} $, where $\Delta$ is the superconducting energy gap. This term reflects errors due to thermal excitations of quasiparticles. The third is two-level systems (TLS) contributions, modelled as $ B T^n $, which represent errors from interactions with two-level fluctuators or coupling to external fields. Here, $ n $ is an exponent dependent on system specifics, and $ B $ is a scaling constant.

For our model, we simplify $ p_{\rm err} $ as follows: the quasiparticle term is negligible at our operating temperatures and is thus omitted. While $ p_0 $ represents the zero-temperature limit, dephasing due to low-temperature environmental fluctuations dominates. References \cite{zhu2024disentangling, de2018suppression, oliver2013materials} suggest that $n = 1$ at temperatures below $\sim 100$ mK and begin to grow quadratically or faster after that. Based on these results, we set $ n = 1 $ and assume the threshold error rate is passed at $\sim 100$ mK. Beyond this temperature our other assumptions break down and the error rate should rise rapidly. This gives $ B = 0.1 $.

\subsection{Numerical Estimates}\label{sub:boundestimates}

Using the transmon-based example from the last subsection, we now perform numerical simulations. We study the system's ability to maintain fault tolerance over time, considering both the presence and absence of cooling mechanisms.

We first present a quasi-linear approximation we used. Simulating the thermal dynamics of large-scale quantum systems over extended periods poses significant computational challenges. In the unbounded-error phase, where the temperature continually increases, we observe three distinct periods in the temperature evolution. First, the temperature rises quickly due to low initial heat capacity. Next, the temperature changes slowly over a prolonged period as heat capacity increases and diffusion equalizes the temperature across the lattice. Finally, the temperature increases rapidly again as the system approaches the error threshold.

The extended slow-change period dominates the simulation time but offers minimal new insights into the system's behaviour. To improve computational efficiency during this phase, we employ a quasi-linear approximation based on the following observations. First, heat diffusion is efficient enough that the lattice temperature becomes nearly uniform, mimicking a scenario where the entire lattice receives heat uniformly. Second, the rate of temperature change becomes approximately constant over time.

\begin{figure}
    \centering
    \includegraphics[width=\columnwidth]{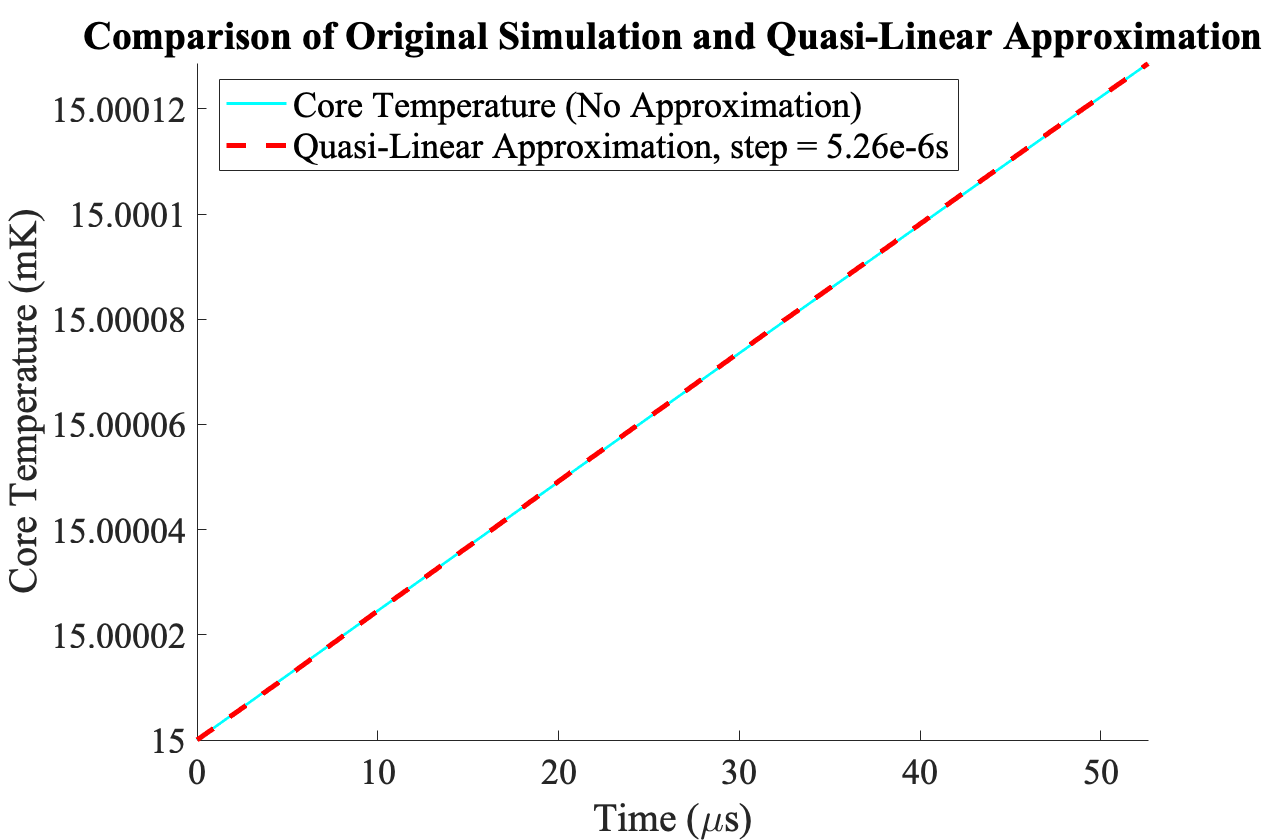}
    \caption{ \caphead{Comparison of Numerical Methods.}
    Temperature evolution obtained from exact numerical simulations (solid line) versus the quasi-linear approximation (dashed line) for a system with $2\cdot10^7$ qubits. The approximation accurately captures the temperature behaviour during the extended slow-change period, reducing computational time without significant loss of accuracy.
    }
    \label{fig:model_simulations}
\end{figure}

Our simulation strategy involves the following. First, we simulate the system until the slope of the temperature-time curve, $T(t)$, stabilizes, indicating a quasi-steady state. Next, we compute the stabilized slope and use it to linearly extrapolate the temperature over fixed longer time intervals. After each extrapolation, we update the temperature grid and recalculate the slope, repeating this process until the temperature begins to increase uncontrollably—typically as the error probability approaches the threshold $p_{\rm th}$. This approach significantly reduces computational time while maintaining accuracy during the less dynamic phase of the simulation. Figure~\ref{fig:model_simulations} compares results from exact numerical simulations and our quasi-linear approximation for a system with $2 \cdot 10^7$ qubits.

\begin{figure}
    \begin{subfigure}[b]{\columnwidth}
        \centering
        \includegraphics[width=\columnwidth]{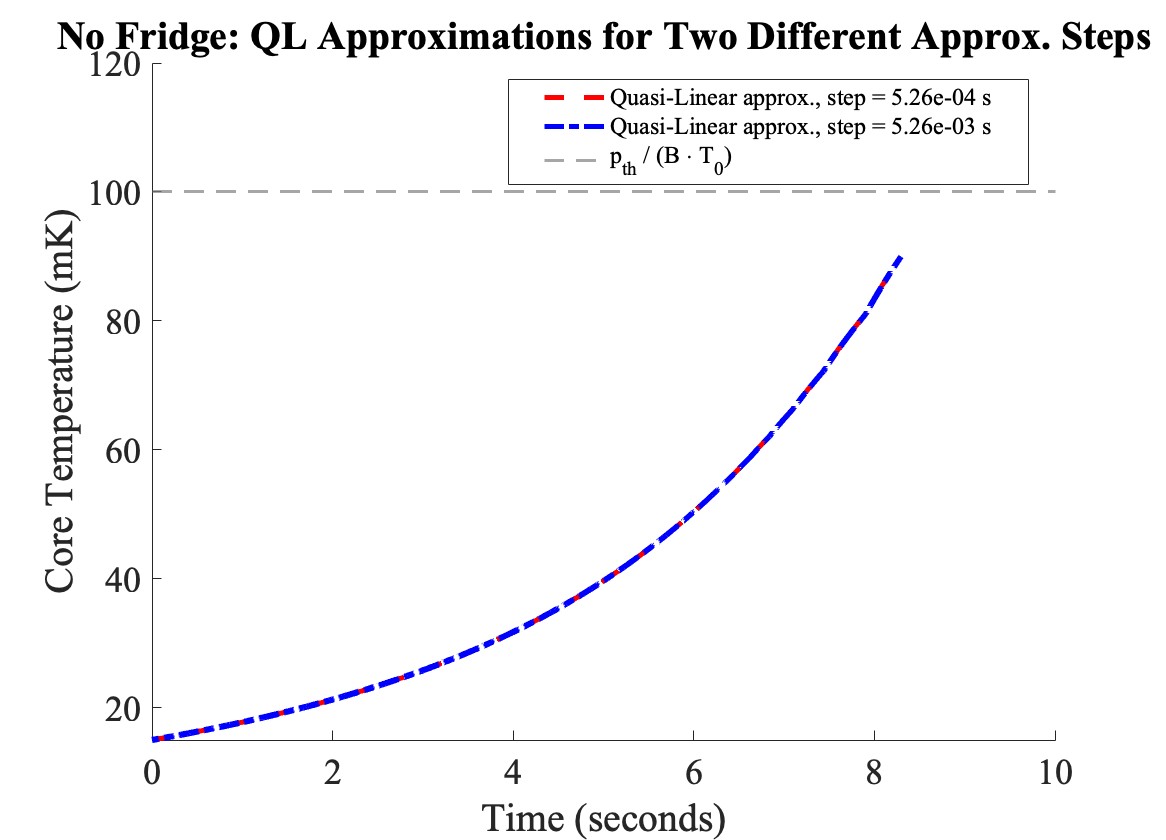}
        \caption{Temperature evolution without cooling.}
        \label{fig:nocooling}
    \end{subfigure}\\
    \begin{subfigure}[b]{\columnwidth}
        \centering
        \includegraphics[width=\columnwidth]{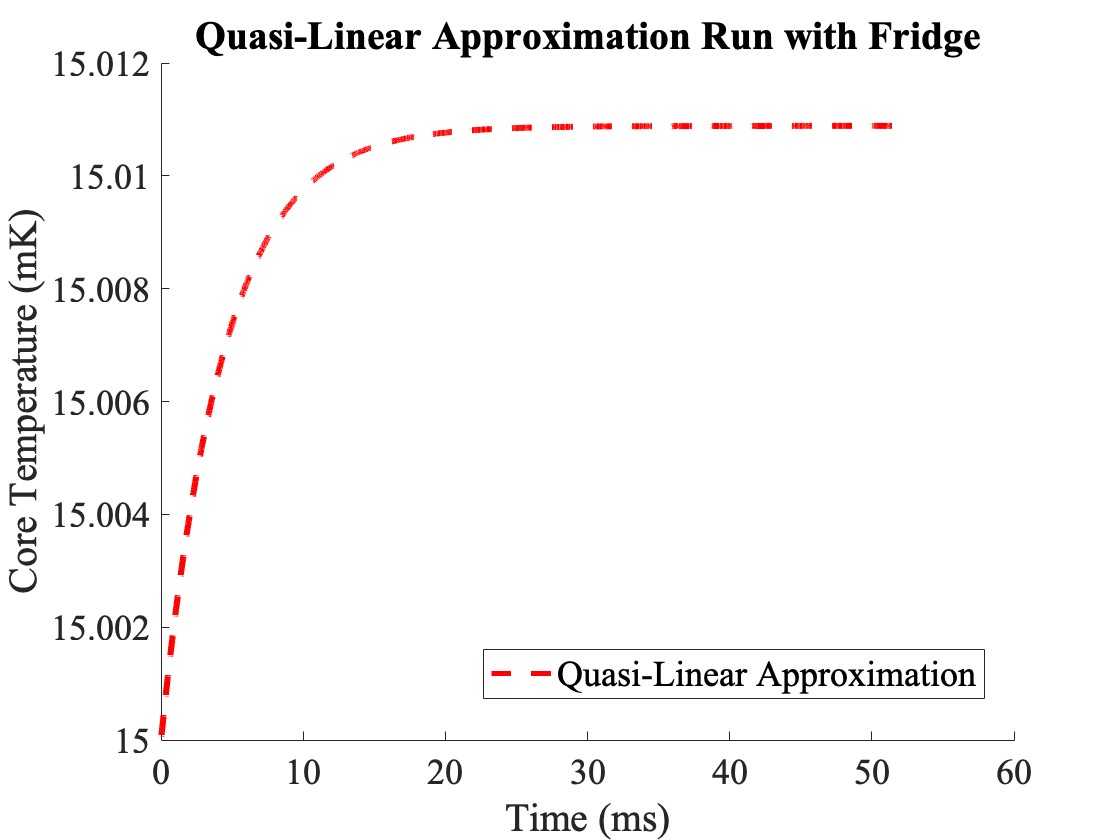}
        \caption{Temperature evolution with cooling.}
        \label{fig:shors_whatregime}
    \end{subfigure}
    \caption{\caphead{Numerical limits using realistic parameters} (a) The temperature immediately surrounding the qubits increases over time without any cooling applied. The system enters the unbounded-error phase, where the temperature continually rises, eventually surpassing the threshold for fault-tolerant quantum computing. (b) Cooling mitigates the temperature increase compared to the no-cooling scenario, preventing the system from entering the unbounded-error phase.}
    \label{fig:numerical_Estimates}
\end{figure}

In the numerics, we first consider a scenario with no cooling ($\gamma = 0$) to establish a baseline for how long computations can proceed before heat accumulation becomes detrimental. Figure~\ref{fig:nocooling} shows the temperature near the qubits as a function of time under these conditions. Since there is no cooling, we are necessarily in the unbounded regime. The computation breaks down on the order of seconds. Next, we include the cooling and compute the computational length possible in Fig.~\ref{fig:shors_whatregime}. We find the temperature stabilizes.

\section{Discussion}\label{sec:dis}

In this work, we explored the thermodynamic limit placed on fault-tolerant quantum computing by the inherent heat dissipation due to quantum error correction (QEC). We constructed a model that allows one to study the competition between the heat generated by QEC and the cooling provided by a fridge. In doing so, we identified a dynamical phase transition in a quantum computer's ability to do fault-tolerant quantum computing. In one phase, the temperature and physical error rate remain bounded, and fault-tolerant quantum computing is possible. In the other phase, the system experiences runaway heating, leading to a diverging error rate that makes fault-tolerant quantum computation impossible. 

One of the motivating questions for this work is: Will Landauer heating, which is unavoidable, stop quantum computers from ever reaching the large scales necessary for tasks like breaking RSA encryption? We used current experimental parameters with our model to produce order-of-magnitude estimates to answer this question. We found that scalable fault tolerance should not be limited by this thermodynamic constraint if current hardware capabilities are maintained as systems scale.

\sg{These considerations are not relevant to near-term quantum devices: in these, the syndromes are stored (and therefore the heating occurs) in classical computers far from the cryostat. However, as platforms scale to millions of physical qubits, it will eventually become necessary to perform tasks like error correction ``on chip,'' leading to Landauer heating. How the generated heat is carried away depends on the platform. In the conceptually simplest case of superconducting qubits, it diffuses to a fridge, leading to the heat-balance dynamics we have considered here. In platforms based on cold atoms or ions, a more natural situation is that the entropy generated by error correction is carried away by radiated photons. In two-dimensional geometries these radiated photons are overwhelmingly likely to leave the system without getting reabsorbed, so the runaway process as we have described it does not occur. Scaling neutral-atom processing to millions of qubits, however, is likely to require modular architectures including elements such as optical cavities~\cite{PRXQuantum.5.020363}. These devices in effect increase the optical depth of the medium and increase the chances that emitted heat will be reabsorbed.
}

Our results open different avenues for future work. One direction is to adapt the model to account for different choices of quantum error-correcting codes, qubit encodings, or noise models and use this extension for a comparative analysis of how these parameters change the thermodynamic limits of fault-tolerant quantum computing. Additionally, while this work focused on the conventional Landauer's bound for heat dissipation, future studies could incorporate the effects of ancillary qubits not being in equilibrium states by utilizing the nonequilibrium quantum Landauer principle~\cite{Goold_2015}. Furthermore, exploring the energetic costs associated with various quantum measurement models~\cite{Abdelkhalek_2018} could further enhance our understanding of the thermodynamics of quantum computation.

More broadly, there are opportunities to investigate quantum effects that could mitigate heat dissipation during QEC. For instance, leveraging symmetries in the system~\cite{majidy2023noncommuting, majidy2024noncommuting, majidy2023non} or engineered dissipation techniques~\cite{harrington2022engineered, shabani2016artificial, pastawski2011quantum} might offer pathways to reduce the thermal burden of error correction. Such advancements could play a useful role in mitigating the thermodynamic challenges identified in this work.

\begin{acknowledgments}
    SM would like to thank Bradley Hauer for many useful discussions and both Chris Pattinson and Bradley Hauer for their initial reading of the manuscript. This work received support from the Banting Postdoctoral Fellowship (SM).
\end{acknowledgments}

\bibliography{refs}
\bibliographystyle{apsrev4-2}

\end{document}